\documentclass[11pt, executivepaper]{article}
\usepackage[utf8]{inputenc}
\usepackage[T1]{fontenc}
\usepackage{natbib}
\usepackage{amsmath}
\usepackage{mathtools}
\usepackage{xcolor}
\usepackage{amsfonts}
\usepackage{graphicx}
\usepackage{enumitem}
\usepackage{geometry}
\usepackage{hyperref}
\hypersetup{colorlinks= true, allcolors=blue}
\setcitestyle{aysep={}}
\begin{document}

\title{\textbf{Observability, Unobservability and the Copenhagen Interpretation in Dirac's Methodology of Physics}}

\author{Andrea Oldofredi\thanks{Contact Information: Universit\'e de Lausanne, Section de Philosophie, 1015 Lausanne, Switzerland. E-mail: Andrea.Oldofredi@unil.ch} \and Michael Esfeld\thanks{Universit\'e de Lausanne, Section de Philosophie, 1015 Lausanne, Switzerland. E-mail: Michael-Andreas.Esfeld@unil.ch}}

\maketitle

\begin{abstract}
Paul A. M. Dirac has been undoubtedly one of the central figures of the last century physics, contributing in several and remarkable ways to the development of Quantum Mechanics (QM); he was also at the centre of an active community of physicists, with whom he had extensive interactions and correspondence. In particular, the British physicist was in close contact with Bohr, Heisenberg and Pauli. For this reason, among others, Dirac is generally considered a supporter of the Copenhagen interpretation of QM. Similarly, he was considered a physicist sympathetic with the positivistic attitude which shaped the development of quantum theory in the twenties. Against this background, the aim of the present essay is twofold: on the one hand, we will argue that, analyzing specific examples taken from Dirac's published works, he can neither be considered a positivist nor a physicist methodologically guided by the observability doctrine. On the other hand, we will try to disentangle Dirac's figure from the mentioned Copenhagen interpretation, since in his long career he employed remarkably different -- and often contradicting -- methodological principles and philosophical perspectives with respect to those followed by the supporters of that interpretation.
\vspace{4mm}

\vspace{4mm}

\center \emph{Preprint of the article published in Quanta, 8:68-87, DOI:10.12743/quanta.v8i1.93}
\end{abstract}
\clearpage

\tableofcontents
\vspace{10mm}

\section{Introduction}

Paul A. M. Dirac has been undoubtedly one of the central figures of the last century physics, contributing in several and remarkable ways to the development of Quantum Mechanics (QM). Many important results bear his name as for instance the bra-ket notation, which became the standard symbolic representation of the mathematical structure of quantum theory, or the well-known equation for the motion of the relativistic electron, which led to the discovery of the existence of anti-matter. 

Moreover, as accurately showed in \cite{Wright:2016}, Dirac was also at the centre of an active community of physicists, with whom he had extensive interactions and correspondence -- contrary to the usual narration of him as a lone genius; in particular, it is interesting to point out for the purposes of the present essay that the British physicist was in close contact with Bohr, Heisenberg and Pauli, who in different ways heavily influenced his work on quantum mechanics. For this reason, among others, Dirac is generally considered a supporter of the so called Copenhagen interpretation, as emphasized for instance in \cite[p. 377]{Bokulich:2004} and by Kragh in his excellent monograph on Dirac's scientific biography \cite[p. 265]{Kragh:1990}. Similarly, he was considered a physicist sympathetic with the positivistic attitude which shaped the development of quantum theory in the twenties, being endorsed among others by notable scientists as Pauli, Heisenberg and the members of the G\"ottingen school who gave decisive contributions to the new quantum physics in those years. 

Against this background, the aim of the present essay is two-fold: on the one hand, we will argue that analyzing specific examples taken from Dirac published works, he can neither be considered a positivist nor a physicist methodologically guided by the observability doctrine, despite the influence that Heisenberg's matrix mechanics -- and the empiricist philosophical principles characterizing it -- exercised on Dirac's approach to quantum theory. On the other hand, we will try to disentangle Dirac's figure from the Copenhagen interpretation of quantum theory, since in his long career he employed remarkably different -- and often contradicting -- methodological principles and philosophical perspectives with respect to those followed by Bohr and Heisenberg.

In more detail, we will discuss two case studies showing that Dirac contradicted the observability doctrine in different ways. For instance, not only he introduced in his physical theories several theoretical entities which were not yet experimentally observed at that time (e.g. the positrons), but also, and more significantly for our purposes, entities which are \emph{in principle} not observable (e.g. the infinite sea of negative energy electrons or his quantum mechanical ether). In this context, the expression ``in principle'' needs some clarification: we are not referring to a \emph{metaphysical} impossibility concerning the ability to observe a particular theoretical entity, nor do we associate such expression with discussions about technological experimental devices employed to measure physical magnitudes. We simply mean that, taking into consideration the mathematical structures of a particular theory, it can be deduced from them that some theoretical entities cannot be observed by construction.\footnote{An alternative, better way to label this impossibility would be \emph{formal} unobservability, which will be used in the remainder of the essay.}  As we will see later on, analyzing these examples it will be clear that in order to construct physical theories, Dirac employed methodological strategies which have been often in conflict with the observability principle.

In the second place, we aim also to underline the several and important differences between Dirac's view concerning the role of the projection postulate in quantum measurements, the correspondence principle and the inter-theoretic relation between classical and quantum theory with respect to those of Bohr and Heisenberg\footnote{To this regard the reader may refer to \cite{Bokulich:2004}.}; consequently, we will argue that it is not completely correct to consider Dirac a fervent supporter of the Copenhagen interpretation of quantum theory. In sum, we claim that there are sound arguments and solid examples taken from his published papers showing that, although his rhetoric was often sympathetic with positivist ideas and/or the Copenhagen interpretation, his work has been in practice much less influenced by these doctrines.
\vspace{2mm}

The structure of the paper is the following: in Section \ref{Obs} we will introduce the observability principle in the context of quantum mechanics taking in particular into account Heisenberg's empiricist attitude at the basis of the matrix mechanics formulation of quantum theory, in Section \ref{3}, we show two case studies in which Dirac explicitly violated the observability doctrine. As mentioned above, we will consider on the one hand, Dirac's interpretation of the sea of negative energy electrons, and on the other, the re-introduction of a quantum mechanical ether in the context of electrodynamics. Section \ref{Cop} will be concerned with the relationship between Dirac and the Copenhagen interpretation of quantum theory. Finally, Section \ref{conc} concludes the paper.

\section{The Observability Principle}
\label{Obs}

Between 1925 and 1927 Heisenberg published two fundamental papers that crucially contributed to the technical development of quantum mechanics\footnote{We are referring here to \cite{Heisenberg:1925, Heisenberg:1927}.}; these essays, in addition, established also a conceptual and philosophical jump with respect to both classical mechanics and the old quantum theory, since matrix mechanics has been formulated taking exclusively observable quantities into account, eliminating concepts referring to unobservable theoretical entities. To this regard, quoting Heisenberg himself, S. Seth writes that 
\begin{quote}
previous approaches to quantum theory could be `seriously criticized on the grounds that they contain, as basic element, relationships between quantities that are apparently unobservable in principle, e.g. position and period of revolution of the electron.' The alternative, he declared, was `to try to establish a theoretical quantum mechanics, analogous to classical mechanics, but in which only relations between observable quantities occur' \citep[p. 840]{Seth:2013aa}. 
\end{quote}

\noindent Interestingly, \citep[p. 19]{Blum:2017aa} note that Heisenberg viewed his ``new theory as merely establishing relations between observable quantities''\footnote{This paper provides an articulated and precise account of Heisenberg's \emph{Umdeutung}, and the authors reconstruct the long and complex process that led Heisenberg to interpret his own theory as a new mechanics, giving the due credit also to the work of Born and Jordan.}, whereas Hilgevoord and Uffink point clearly out that Heisenberg's leading idea, in his founding papers on quantum theory, was that only quantities in principle observable and/or measurable should appear in the vocabulary of a physical theory, ``and that all attempts to form a picture of what goes on inside the atom should be avoided. In atomic physics the observational data were obtained from spectroscopy and associated with atomic transitions. Thus, Heisenberg was led to consider the ``transition quantities'' as the basic ingredients of the theory'' \citep[Section 2.1]{Uffink:2016}. 

In general terms, it is possible to claim that in those years Heisenberg found it methodologically unsatisfactory to formulate a mechanical theory in order to explain the new observed phenomena making use of notions as position and orbit of physical objects -- which were not observable --, given that theoretical frameworks based on such concepts were shown to be empirically inadequate. Then, he provided a new theory establishing rules for the observed transition amplitudes and frequencies, taking the latter as the fundamental objects of his theory, and consequently, every physical quantity in that framework could have been associated to something observable. Successively, Born and Jordan interpreted Heisenberg's theory as a proper mechanical theory originating what has become known as matrix mechanics, the first proper formulation of quantum mechanics \citep[Section 7]{Blum:2017aa}.
In addition, in order to understand Heisenberg's empiricist views, it is worth taking into account the developments of the conceptual shift contained in \cite{Heisenberg:1927}, the essay in which the first formulations of the uncertainty relations were given. In this paper, Heisenberg employed explicit operational assumptions associating the meaning of concepts and notions of his theory -- for instance the position or the momentum of a quantum particle -- to experimental procedures capable of measuring them. Hence, he provided the uncertainty principle with an ontological interpretation according to which quantum particles cannot have simultaneously well-defined values for position and momentum.\footnote{To this regard the reader may refer to a note added to \cite{Heisenberg:1927}; \cite{Uffink:2016} provide a precise reconstruction of Heisenberg's interpretation of the uncertainty principle.}

Let us call this positivistic attitude the \emph{ontological observability principle}. From a philosophical perspective, the positive content of this principle implies that a meaningful physical theory should be exclusively and uniquely about what can be experimentally observed and/or manipulated. Alternatively stated, the negative content of such principle states that objects, quantities or processes which cannot be directly or indirectly observed should not be part of the vocabulary of any theoretical building, therefore, they should not be considered meaningful terms of scientific theories. Since the latter guide our \emph{Weltanschauung}, i.e. our interpretation of the physical reality, and from them we should reconstruct our manifest image of the world, it follows that unobservable theoretical entities should be excluded from any ontological commitment. Heisenberg, however, was not the only supporter of the positivist doctrine, as clearly stated by \cite[p. 262]{Kragh:1990}
\begin{quote}
in the context of quantum physics the observability doctrine is often referred as to Heisenberg's observability principle, although Heisenberg in fact used it after Pauli, who stated it clearly in 1919: ``However, one would like to insist that only quantities which are in principle observable should be introduced in physics''. [...] Quantities that were in principle unobservable would be according to Pauli, ``fictitious and without physical meaning''.
\end{quote}
 
According to Pauli's perspective, observability is tightly connected with measurements\footnote{To this regard, Wolff interestingly points out that the notion of observability is untenable if it is associated with mere perceptibility, for a detailed argument the reader may refer to \cite{Wolff:2014}.}, in the sense that what cannot be measured (directly or indirectly) should not enter into physical theories. This statement clearly entails the normative ontological import of the observability principle mentioned above: if some object, process or property cannot be in principle or in practice, directly or indirectly observed, manipulated or measured, then it should be considered a fictitious entity, meaningless and consequently not referring to anything. 

A \emph{locus classicus} widely cited to expose Pauli's ideas concerning the observability doctrine can be found in his critique of Weyl's unified theory (appeared in \cite{Weyl:1920}). Pauli was strongly opposed to the idea of an electric field which was \emph{inside or internal} to the electron. His argument goes as follows: the field strength can only be defined as a force on a certain test-body. Given that the smallest test-body available is the electron itself, it follows that one cannot introduce the field strength into a mathematical point, for it would be a quantity in principle non-measurable since every possible measurement would be inappropriate lacking a test-body. Furthermore, as Johanna Wolff recently pointed out, non-measurable quantities threaten the empirical testability of physical theories: ``If a theory stipulates that a certain relationship holds between several quantities, or that changes in a particular quantity are responsible for certain phenomena, then it is desirable to have some measurement procedure, direct or indirect, of that quantity in order to be able to generate specific predictions and to test them. If there are reasons to believe, not only that such measurements are unavailable, but that they are impossible, then this counts against the theory in question'' (\cite{Wolff:2014}, p. 21). Since the electric field inside an electron cannot be in principle measured, this very notion has no physical meaning. Therefore, one should consider it as referring to anything. Apart from Pauli and Heisenberg, such an observability doctrine was also shared by the people in the G\"ottingen school, as declared by Heisenberg himself in the fifth session of his \href{http://aip.org/history-programs/niels-bohr-library/oral-histories/4661-1}{interview with T. Kuhn}\footnote{The reader may refer to ``Interview of Werner Heisenberg by Thomas S. Kuhn and John Heilbron on 1962 November 30, Niels Bohr Library $\&$ Archives, American Institute of Physics, College Park, MD USA''.}:
\begin{quote}
[w]hen one spoke about special relativity, people always said, ``Well, there was this very famous point of Einstein that one should only speak about those things which one can observe, that actually the time entering in the Lorentz transformation was the real time.'' And in some way that was an essential turn which Einstein had given to the Lorentz idea. Lorentz had the right formulas, but he thought that was the apparent time. Einstein said, however, ``There is no apparent and no real time; there is just one real time, and that is what you call the apparent time.'' So this turning of the picture by saying the real things are those which you observe and everything else is nothing was in the minds of the G\"ottingen people.
\end{quote}

Against this background, and alongside the ontological observability principle considered so far, it is worth mentioning what can be called the \emph{causal observability principle}, which is contained in \cite{Wolff:2014}. According to Wolff, it is not correct to claim that the notion of observability employed by \cite{Heisenberg:1925} is concerned with measurements, but rather, it should be led back to the notion of causal inefficacy. More precisely, the notions of electron's orbit and electron's position are not meaningless in virtue of the absence of hypothetical or actual experimental procedures able to measure such quantities. Instead, the latter do not play any causal, mechanical role in the theory. Therefore, employing a sort of Ockham razor's, Heisenberg eliminated such concepts from his own matrix mechanics. In this manner, Wolff promotes a view of the observability principle in which ontological commitments are disentangled from measurability conditions. 

In this essay our aim is not to take position for a particular interpretation of the observability doctrine in the context of Heisenberg exegesis; it is however interesting to take into account different versions of the observability principle, since we are going to show that in his works Dirac introduced several theoretical entities which are not only in principle unobservable, but also causally inefficient. Thus, Dirac's theories violate the observability doctrine in both these variants. 

Interestingly, there are several sources in which Dirac explicitly indicated Heisenberg's influence on his own work on relativistic quantum theory and the methodological relevancy of the observability principle. In particular, Dirac thought that Heisenberg's departure from classical theory was the recognition of the non-commutative nature of the algebra of quantum observables: ``that one should confine one's attention to observable quantities, and set up an algebraic scheme in which only these observable quantities appear'' \citep{Dirac:1932}. Another source worth mentioning is the first session of \href{http://aip.org/history-programs/niels-bohr-library/oral-histories/4575-3}{Dirac's interview with Kuhn and Wigner}\footnote{The reader may refer to ``Interview of P. A. M. Dirac by Thomas S. Kuhn on 1963 May 7, Niels Bohr Library $\&$ Archives, American Institute of Physics, College Park, MD USA''.}:

\begin{quote}
Heisenberg made his trip to Cambridge, in I think, June, or it might have been July, of 1925. He gave a talk about a new theory to the Kapitza Club, but I wasn't a member of the club so I did not go to the talk. I did not know about it at the time. The first I heard of it was in September when Fowler sent to me a copy of the proofs of Heisenberg's paper [Dirac is referring to \cite{Heisenberg:1925}] and asked me what I thought about it. That was the first that I heard about it. I think Fowler found it interesting. He was a bit uncertain about it and wanted to know what my reaction to it would be. When I first read it I did not appreciate it. I thought there wasn't much in it and I put it aside for a week or so. Then I went back to it later, and suddenly it became clear to me that it was the real thing. And I worked on it intensively starting from September 1925. I think it is just a matter of weeks or so before I got this idea of the Poisson brackets.
\end{quote}
\noindent Moreover, he added that in discovering a better quantum theory able to overcome the Bohr-Sommerfeld quantization method ``Heisenberg's idea provided the key to the whole mystery'', with the introduction of matrix mechanics.

In addition, there are many other examples showing the influence of the observability doctrine on Dirac, since he considered neutrinos to be unobservable and therefore without physical existence, similarly he excluded from the set of existent objects the interior of black holes.\footnote{Interestingly, in both cases Dirac made patently wrong statements about these unobservable entities, since, on the one hand, neutrino have been experimentally observed in the mid-Fifties and classified among the group of leptons of the standard model of particle physics, on the other, Hawking showed that black holes can emit radiation, and thus, they are not in principle unobservable entities.} As quoted by \citep[p. 80]{Kragh:1990}, Dirac stated in 1927 that 
\begin{quote}
The main feature of the new theory [i.e. quantum mechanics] is that it deals essentially only with observable quantities, a very satisfactory feature. One may introduce auxiliary quantities not directly observable for the purpose of mathematical calculation; but variables not observable should not be introduced merely because they are required for the description of the phenomena according to ordinary classical notions [...] The theory enables one to calculate only observable quantities [...] and any theories which try to give a more detailed description of the phenomena are useless.
\end{quote}

In sum, the influence of positivistic ideas and the observability doctrine endorsed by Heisenberg and Pauli can be traced in many places in Dirac's writings. Nonetheless, it is legitimate to ask whether or not such influence really had a decisive impact on Dirac's methodology of physics. In what follows we will answer this question in the negative, explaining that Dirac cannot be considered a physicist with a positivist attitude, nor that the observability principle played a crucial role in his works on quantum mechanics. 

\section{Unobservable Theoretical Entities in Dirac's Work}
\label{3}

In this section we aim to argue that Dirac should be neither considered a supporter of a positivist philosophy, nor a physicist methodologically guided by the observability doctrine. In order to support our thesis, in fact, we are going to consider two case studies taken from his work in which he explicitly postulated theoretical entities which are \emph{formally not observable} in the sense defined in the first section. In our view these examples show that, although Dirac rhetorically embraced the observability principle in many places, in practice it played a considerably little role in his methodology. 

In what follows, we consider in the first place how Dirac arrived to interpret the negative energy solutions of his relativistic equation for the electron as positrons, whereas in the second subsection, we discuss the re-introduction of the ether notion in the context of quantum electrodynamics.\footnote{Although Dirac was not constrained in formulating new theories or advancing new ideas by the observability principle, he cannot be considered a scientific realist in the usual sense. To this regard, in fact, many scholars argued that Dirac's realism is tightly connected with -- and guided by -- aesthetic considerations concerning mathematical beauty and elegance of a given physical theory. For lack of space we cannot discuss this issue in what follows. For details the reader may refer to \cite{Bueno:2005}, \cite{Kragh:1990}, \cite{Pashby:2012}, \cite{Wright:2016} among others.}

\subsection{Case study I: The Dirac Equation}

Let us then consider our first case study, the Dirac relativistic quantum mechanical equation for the electron, written for the first time by Dirac in his well-known 1928 paper ``\emph{The quantum theory of the electron}'' \citep{Dirac:1928}. This equation is a milestone in the physical literature, being the first result able to combine and satisfy the axioms of both quantum mechanics and special relativity theory. Furthermore, from this equation it has been possible to predict the existence of anti-matter. Indeed, we will here cover the main steps that led Dirac to the interpretation of the negative energy solutions of his equation as positrons, the anti-particles of the electrons.

In order to provide an elementary introduction to this equation, we start from non-relativistic quantum mechanics, where the total energy of physical systems is defined as the sum of kinetic and potential energy:
\begin{align}
\label{en}
E=T+V=\frac{p^2}{2m}+V 
\end{align}
\noindent{then}, taking into account usual quantization procedures, one defines energy and momentum operators: 
$$\textbf{p}\longrightarrow{-i\hbar\nabla}, \nonumber$$

$$E\longrightarrow{i\hbar\frac{\partial}{\partial{t}}}. \nonumber$$

From the definitions of the energy on the one hand, and the energy and momentum operators on the other, it is straightforward to write a single-particle Schr\"odinger Equation (SE), which is the fundamental dynamical law of quantum theory.\footnote{We recall that the Schr\"odinger equation is not strictly derivable in quantum mechanics; thus, the reader must consider what we have written as a simple and heuristic view to easily introduce such law starting from the definition of the energy, since it will be crucial for the introduction of negative energy solutions in the context of the Dirac equation.} To obtain such result, it is sufficient to multiply each side of (\ref{en}) with the wave function of a particle $\psi(x,t)$:
\begin{align*}
E\psi(x, t)=(T+V)\psi(x, t) 
\end{align*}  
\noindent{With} obvious substitutions, we arrive at the usual form: 
\begin{align}
\label{SE}
i\hbar\frac{\partial}{\partial t}\psi(x,t)=-\frac{\hbar^2}{2m}\Delta\psi(x, t) +V\psi(x, t) =\hat{H}\psi(x,t)
\end{align}
\noindent where $\hbar$ is the Planck constant and $\hat{H}$ is the usual Schr\"odinger Hamiltonian, i.e. the sum of kinetic and potential energy of the system at hand.

However, SE is not a reliable description of particles' dynamics if one takes into consideration relativistic quantum mechanics, since space and time derivatives are not of the same order, i.e. are not treated symmetrically as required by relativity.\footnote{In the Schr\"odinger equation the derivative of the time coordinate is first order, whereas the derivative of spatial coordinates is second order.} In order to make the Schr\"odinger equation relativistic, one may start from the relativistic energy-momentum relation:
\begin{align*}
E^2=\textbf{p}^2c^2+m^2c^4. 
\end{align*}

\noindent Following the straightforward procedure used to obtain \eqref{SE} a few lines above, but making now use of the relativistic energy-momentum relation, one obtains the well-known Klein-Gordon equation (KGE) via simple substitutions:
\begin{align}
\label{KGE}
-\hbar^2\frac{\partial^2}{\partial{t^2}}\psi(x, t)=-\hbar^2c^2\nabla^2\psi(x, t)+m^2c^4\psi(x, t). 
\end{align}

\noindent Contrary to the case of the SE, the KGE is a second order equation in both spatial and temporal coordinates. Being Lorentz covariant, KGE satisfies the symmetry between space and time coordinates imposed by relativity. This equation has plane waves solutions of the form $\psi=Ne^{\pm{i}(\textbf{p}\cdot{\textbf{r}}-Et)}$, hence, it yields also \emph{negative} energies solutions as consequence of the relativistic energy-momentum relation\footnote{More specifically, for such solutions the KGE gives $-E^2\psi=-|\textbf{p}|^2\psi-m^2\psi$ which entails $E=\pm\sqrt{|\textbf{p}|^2+m^2}$, so that one obtains straightforwardly also negative solutions.}:
\begin{align*}
E=\sqrt{\textbf{p}^2c^2+m^2c^4}\quad \text{and} \quad E=-\sqrt{\textbf{p}^2c^2+m^2c^4}.
\end{align*}
When this equation appeared in 1926, it was unclear how to physically interpret the meaning of the negative energy solutions, as well as the negative probability densities associated with them, since the wave function would not anymore be interpretable as a probability amplitude in the context of the KGE. In the second place, this equation did (and does) not allow for a treatment of particles with spin. Thus, due to these difficulties, the KGE could not been considered a successful generalization of \eqref{SE}.\footnote{Equation \eqref{KGE} is usually presented in this form: $$\frac{1}{c^2}\frac{\partial^2}{\partial{t^2}}\psi-\nabla^2\psi + \frac{m^2c^2}{\hbar^2}\psi=0,$$ which is equivalent to the formulation written above. This equation was first discovered by Schr\"odinger who, however, discarded it for the already mentioned difficulties. The KGE is now a well-established result in quantum field theory, since it describes spin-0 particle fields, among which we find the Higgs boson.}

These were the principal motivations which led Dirac to propose a new, different equation of motion for the relativistic electron which could properly be considered an extension of the Schr\"odinger equation, the well-known Dirac Equation (DE)\footnote{To maintain a coherent notation in the essay, we will avoid to make use of the covariant version of DE.}: 
\begin{align}
\label{Dirac}
i\hbar\frac{\partial}{\partial{t}}\psi(x, t)=\Bigg(\beta{m}c^2+c\bigg(\sum_{n=1}^3\alpha_n p_n\bigg)\Bigg)\psi(x, t)\equiv\hat{H}_{Dirac}\psi(x, t),
\end{align}
\noindent which is a first order equation in both the spatial and the temporal coordinates in order to preserve Lorentz invariance. Here the wave function represents an electron of mass $m$ and spacetime coordinates $(x, t)$, $c$ is the speed of light, and the l.h.s. of this equation represents the Dirac Hamiltonian. The latter contains the terms $\alpha$, $\beta$ which are $4\times4$ matrices, where $\beta^2=\alpha^2_x=\alpha^2_y=\alpha^2_z=I_4$, $\beta\alpha_j+\alpha_j\beta=\alpha_j\alpha_k+\alpha_k\alpha_j=0$, for every $j\neq{k}$; therefore, $\alpha, \beta$ are Hermitian, their square is equal to the identity matrix, and finally, they anti-commute. The usual representations of these matrices are:

\begin{align}
\alpha_n=
   \begin{pmatrix}
      0 & \sigma_i \\
      \sigma_i & 0 \\ 
   \end{pmatrix} \nonumber
\end{align}

\noindent{and}
\begin{align}
\beta=
   \begin{pmatrix}
      1 & 0\\
      0 & -1 
   \end{pmatrix} \nonumber
\end{align}
\noindent{where} the $\sigma_i$ are the usual Pauli's matrices.\footnote{In brief, the Pauli matrices are three $2\times2$ complex, Hermitian and unitary matrices of the form: $$\sigma_x=
\begin{pmatrix}
0 & 1\\
1 & 0
\end{pmatrix}, \quad \sigma_y=
\begin{pmatrix}
0 & -i\\
i & 0
\end{pmatrix}, \quad  \sigma_z=
\begin{pmatrix}
1 & 0\\
0 & -1
\end{pmatrix}.$$
These objects have been introduced by Pauli as the observables corresponding to the particles' spin along the axis of the three-dimensional Euclidean space. For a detailed analysis of the Dirac equation the reader may refer to \cite{Thaller:1992aa}, Chapter 1.} 

Since the $\alpha_i,\beta$ terms appearing within the Dirac Hamiltonian are $4\times4$-matrices as stated above, the wave function which is a solution of this equation must be a $4-$component object which in the physical literature is called \emph{Dirac spinor}: 
\begin{align}
\psi=
\begin{pmatrix}
\psi_1\\
\psi_2\\
\psi_3\\
\psi_4
\end{pmatrix}. \nonumber
\end{align}
These solutions of the Dirac equation, these four component spinors, naturally include the treatment of the particles spin, which was not treated by the Klein-Gordon equation, in virtue of the presence of the Pauli matrices. Moreover, contrary to the case of \eqref{KGE}, the question of the probability density is solved in the case of the DE, since it is always positively definite (cfr. \cite{Thaller:1992aa}, Chapter 1).

Nevertheless, the crucial problem concerning a consistent physical interpretation of the negative energies remained. In order to analyze this issue, it may be useful to introduce the solutions of the Dirac equation for a free particle at rest ($\textbf{p}=0$)\footnote{In general, wave functions of the form  $\psi=u(E, \textbf{p})e^{i(\textbf{p}\cdot\textbf{r}-Et)}$ are solutions of the DE, where $u(E, \textbf{p})$ is a constant four-vector component spinor.}:
\begin{align*}
\psi=u(E,0)e^{-iEt} 
\end{align*}
\noindent{where} $u(E,0)$ is a constant four component spinor which satisfies the DE. The Dirac equation has four mutually orthogonal solutions of the form\footnote{Here we are including the time dependence from $\psi=u(E, 0)e^{-iEt}$.}:
\begin{align*}
u_1(m,0)= \psi_1=
\begin{pmatrix} 
1\\
0\\
0\\
0
\end{pmatrix};\quad
u_2(m,0)= \psi_2=
\begin{pmatrix}
0\\
1\\
0\\
0
\end{pmatrix};
\end{align*}
\begin{align*}
u_3(m,0)=\psi_3=
\begin{pmatrix}
0\\
0\\
1\\
0
\end{pmatrix};\quad
u_4(m,0)=\psi_4=
\begin{pmatrix}
0\\
0\\
0\\
1
\end{pmatrix}. 
\end{align*}

Looking at these solutions, it is straightforward to note that there are two spin states with positive energy ($E>0$), $\psi_1$ and $\psi_2$, and two spin states with negative energies ($E<0$), $\psi_3$ and $\psi_4$. It must be said, furthermore, that the latter two components cannot be simply discarded and considered surplus mathematical structure without any physical meaning, since in quantum mechanics one needs a \emph{complete} set of states for the energy, i.e. all the four solutions in this case. Another problem related to such negative energy solutions concerned the physical explanation of the stability of positively charged electrons; a crucial question to answer was, in fact, why they do not fall into lower and lower energy states. To solve these issues Dirac provided several possible interpretations for these negative energy solutions, arriving at the beginning of the thirties -- precisely in 1931 -- at the well-known Dirac sea, or hole theory, which led to the discovery of the first anti-particle, the positron. 

Through a brief analysis of the main steps followed by Dirac to arrive at this theory, we will see that he employed methodological criteria in contradiction with the observability principle in order to provide the DE with a consistent solution to the $\pm e$ difficulty, as physicists used to call it.
\vspace{2mm}

In the first place, let us point out that the problem of negative energy solutions of the DE was strictly related to both philosophical and physical issues: at the end of the twenties atomic physics included only two kinds of particles in its ontology, the positive energy electron and the proton; moreover, since negative energy solutions were generally considered \emph{unphysical}, such negative energy electrons were considered at those times non-existent entities. Nonetheless, as stated above, in the case of the DE it was not possible to simply discard them, consequently, physicists had to find a sound interpretation for such states. This question was particularly urgent for Dirac, since he was well aware that perturbations could cause transitions from states with positive energy to states of negative energy: ``[s]uch a transition -- claimed Dirac -- would appear experimentally as the electron suddenly changing its charge from $-e$ to $e$, a phenomenon which has not yet been observed'' \citep[p. 612]{Dirac:1928}. Other physicists like Jordan, Klein, Pauli and Heisenberg found the $\pm$ difficulty a serious trouble for Dirac's relativistic theory of the electron; in particular Klein showed that the motion of a simple electron moving against a barrier is not correctly described by the DE, since it provides results completely different from those of the Schr\"odinger equation. Similarly, another negative consequence entailed by the expression $E=-\sqrt{\textbf{p}^2c^2+m^2c^4}$ was that the energy of the electrons decreased the faster they moved. 

To overcome these puzzling issues, in between 1928 and 1929\footnote{It is important to recall that the Dirac sea hypothesis did not appear in \cite{Dirac:1928}, on the contrary, in that essay he wrote that negative solution could have been simply discarded (cf. \cite{Dirac:1928}, p. 618).} Dirac proposed a solution of the $\pm e$ difficulty interpreting the vacuum as filled with an infinite density of negative energy electrons, regarding the vacancies or ``holes'' in such sea as \emph{protons}.\footnote{For a detailed historical reconstruction of the proton interpretation of holes, the reader may refer to \cite{Kragh:1990}, Chapter 5.} A similar idea was advanced by Weyl in the spring of 1929, however, he identified the negative electrons with protons. Dirac knew Weyl's proposal but was not convinced by the interpretation of the sea of negative protons for three main reasons \citep[p. 362]{Dirac:1930}:
\begin{enumerate}
\item It would entail a violation of the law of electric charge conservation: if there would be a transition of an electron from a positive to a negative state, we would assist to a transition from an electron to a proton with the consequent charge difference.
\item A negative energy electron would have less energy the faster it moves, so that it must absorb energy in order to be at rest, but Dirac stated that ``[n]o particle of this nature have ever been observed''.
\item If a negative energy electron would be a proton, it would simultaneously repelled and attracted by positive energy electrons.
\end{enumerate}

\noindent Against this background, Dirac's own view has been expressed very clearly in a letter sent to Bohr on 29th November 1929:
\begin{quote}
There is a simple way of avoiding the difficulty of electrons having negative kinetic energy. Let us suppose the wave equation
\begin{align*}
\bigg[ \frac{W}{c}+\frac{e}{c}A_0+\rho_1(\vec{\sigma}\cdot\vec{\gamma}+\frac{e}{c}A)+\rho_0mc\bigg]\psi=0
\end{align*}
\noindent does accurately describe the motion of a single electron. Let us now suppose there are so many electrons in the world that all these most stable states are occupied. The Pauli principle will then compel some electrons to remain in less stable states. For example if all the states of $-$ve energy are occupied and also few of $+$ve energy, those electrons with $+$ve energy will be unable to make transitions to states of $-$ve energy and will therefore have to behave
quite properly. The distribution of $-$ve electrons will, of course, be of infinite density, but it will be quite uniform so that it will not produce any electromagnetic field and one would not expect to be able to observe it. It seems reasonable to assume that not all the states of negative energy are occupied, but that there are a few vacancies or ``holes''. Such a hole which can be described by a wave function like an X-ray orbit would appear experimentally as a thing with $+$ve energy, since to make the hole disappear (i.e. to fill it up,) one would have to put $-$ve energy into it. Further, one can easily see that such a hole would move- in an electromagnetic field as though it had a $+$ve charge. These holes I believe to be the protons. When an electron of $+$ve energy drops into a hole and fills it up, we have an electron and proton disappearing simultaneously and emitting their energy in the form of radiation. (Quoted in \cite{Wright:2016}, pp. 232-233).
\end{quote}

Successively, in the essay \emph{A Theory of Electrons and Protons} published in 1930, Dirac describes in detail this idea. Following the argumentative lines of the letter quoted above, he started from the consideration that the most stable states for the electrons -- the states with lowest energy -- were those with negative energy and high velocity, then, he continued, all the electrons will tend to fall into such states emitting radiation. The Pauli exclusion principle, however, will prevent ``more than one electron going into any one state''. Therefore, Dirac proposed to consider the overwhelming majority of the infinite density of negative energy stets as occupied. Interestingly, such infinity is \emph{actual} in Dirac's theory, since he explicitly stated that ``[w]e shall have an infinite number of electrons in negative-energy states, and indeed an infinite number per unit volume all over the world, but if their distribution is exactly uniform we should expect them to be completely unobservable'' \citep[p. 362]{Dirac:1930}. According to the principal idea of this essay, the holes in this infinite sea of negative energy electrons are protons: 
\begin{quote}
[w]e are therefore led to the assumption that \emph{the holes in the distribution of negative- energy electrons are the protons}. When an electron of positive energy drops into a hole and fills it up, we have an electron and proton disappearing together with emission of radiation \citep[p. 363]{Dirac:1930}.
\end{quote}

Concerning this particular solution to the $\pm e$ difficulty, Kragh stressed that the methodological principle guiding Dirac was a belief about the unity of nature, i.e, the idea for which matter was composed by two essential ingredients, electrons and protons -- neutrino have been recently proposed at the time but physicists were skeptical about their actual existence, and the neutron was then referred to the electron-proton pair, and not regarded as a new kind of particle. Hence, following the available knowledge in atomic physics and its practice, he preferred not to introduce new theoretical entities -- to this specific regard Kragh speaks also about sociological concerns given the very conservative attitude toward the proliferation of new particles in the physicists community. Be that as it may, Dirac had no problem in introducing a formally unobservable sea of negative electrons in his theory in order to solve the $\pm e$ difficulty, a move that would not be justified according to the observability doctrine. Another instantiation of the unity of nature principle is traceable in those years, since Dirac conceived  the possibility to reduce the then known species of elementary particles just to the electron. More precisely, he thought that electrons and protons were not independent objects but different manifestations of the very same fundamental particle:
\begin{quote}
It has always been the dream of philosophers to have all matter built up from one fundamental kind of particle, so that it is not altogether satisfactory to have two in our theory, the electron and the proton. There are, however, reasons for believing that the electron and proton are really not independent, but are just two manifestations of one elementary kind of particle \citep[p. 605]{Dirac:1930b}. 
\end{quote}

Despite Dirac's efforts in trying to provide a physically sound and metaphysically elegant solution to the $\pm e$ difficulty, his interpretation of holes as protons did not convinced many physicists. In particular, Heisenberg calculated the  electron-proton interaction according to Dirac's theory arriving to the conclusion that these two particles would have the very \emph{same} mass, a result which is contradicted by already available evidence, protons having a much heavier mass with respect to electrons. 

Successively, confronted with Heisenberg's objection, and with many other physical, mathematical and philosophical critiques to the proton hypothesis\footnote{Dirac himself was aware of the inherent difficulties of this proposal, but the force of the unitary idea of matter was superior, as noted in \citep[pp. 96-99]{Kragh:1990}. Interestingly, Dirac was not much impressed by empirical arguments (for instance those provided by Oppenheimer among others) showing contradictions with the available evidence, but he readily detached himself from this hypothesis once formal and rigorous proofs from Pauli, Heisenberg and Weyl were advanced. Here we can easily see how the force of mathematical reasoning had a greater impact on Dirac in comparison to the empirical evidence and knowledge.}, Dirac abandoned it, turning to the \emph{positive electron} idea. He introduced for the first time in \cite{Dirac:1931} a hypothetical particle with the same mass of the electron but with opposite charge\footnote{Ironically, Dirac thought that also the proton could have its unfilled hole so that ``[i]n a few lines he doubled the number of elementary particles'' \citep[p. 104]{Kragh:1990}.}:
\begin{quote}
A hole, if there were one, would be a new kind of particle, unknown to experimental physics, having the same mass and opposite charge to an electron. We may call such a particle an anti-electron \citep[p. 61]{Dirac:1931}.
\end{quote}

In his 1931 paper, Dirac proposed the well-known revolutionary hypothesis about the nature of the vacuum, suggesting to consider it as composed by an infinity of negative energy electron eigenstates, specifying this time that \emph{all} -- ``and not nearly all'' \citep[p. 61]{Dirac:1931} -- such states are occupied. The holes, then, were interpreted as positrons, and this idea became the standard interpretation of such negative energy solutions of the DE. To this regard, Bohm and Hiley\footnote{It is interesting to underline that the Dirac sea has been reformulated in the context of Bohmian mechanics. On the one hand, Bohm himself in \cite{Bohm:1953aa} and \cite{Bohm:1993aa} Chapter 12 discussed in depth this hypothesis in order to extend the pilot-wave approach to the treatment of the fermionic field. On the other hand, contemporary mathematical physicists are working on the developments of the Dirac sea approach, the reader interests in this approach may  concentrate especially on  \cite{Colin:2003aa}, \cite{Colin:2003ab} and \cite{Colin:2007aa}. For a more formal treatment the reader should refer to \cite{Deckert:2010aa}, \cite{Deckert:2015aa} and \cite{Deckert2010a}. A philosophical discussion of the Dirac sea in Bohmian mechanics is contained in \cite{Deckert:2019}.} summed up the hole theory in a concise and precise manner, characterizing it as follows:
\begin{quote}
What he [Dirac] proposed was that in the vacuum, all the negative energy states were filled. Because the particles satisfy the exclusion principle, the transition of particles to the negative energy states could therefore never occur. However, when a particle went to a positive state, it would leave a hole that acted like an antiparticle. So in effect a positive and a negative pair would be created in such process \citep[p. 276]{Bohm:1993aa}.
\end{quote}
\vspace{2mm}

Having roughly presented the crucial steps that led Dirac to postulate the positron hypothesis, let us now consider some philosophical and methodological aspects of the above discussion. 
In the first place, it is straightforward to understand that Dirac violated the observability principle in two different ways with the hole theory: on the one hand, positrons were not yet observed quantities -- they would have been discovered by Carl Anderson at Caltech only in 1932 --, contrary to the case, for instance, of transition amplitude in Heisenberg's matrix mechanics, on the other hand and more importantly, the sea of negative energy electrons is formally unobservable. According to the Dirac sea picture, as already pointed out, the holes are the only observable entities, whereas the sea must be unobservable since \emph{all} the negative energy states are occupied in virtue of the Pauli exclusion principle; therefore, it follows that being electrons in negative states in a perfect homogeneous distribution, it is impossible to interact or observe such sea. Moreover, we do not have empirical evidence for such infinite density of negative energy electrons in the actual world, so that the sea is also causally inert.\footnote{This feature is also a logical consequence of the Pauli exclusion principle, since negative energy electrons are homogeneously distributed and do not take part in interactions.} Hence, in order to provide the $\pm e$ difficulty a sound physical solution, Dirac did not hesitate to make abstract and audacious speculations involving unobservable entities, showing in practice how the observability doctrine has not been an interesting methodological option in formulating the hole theory.\footnote{It must be noted that many physicists raised strong objections also to Dirac's theory as proposed in 1931: some viewed the infinite sea of positrons as an entity close to the classic ether, regarding it as a purely metaphysical notion, as communicated to Dirac by Igor Tamm in a letter dated 5th June 1933. As reported also by Kragh, notable figures like Landau and Peierls evaluated negatively Dirac's theory affirming that it was ``senseless'' in virtue of the presence of this unobservable sea. Similarly, Bohr was unhappy with the hole theory and he expressed all his perplexities in many letters to Dirac, most of them related with the infinite negative -- but not observable -- electric charged introduced with the positron sea. Strong opposition came also from Pauli who refused to associate the new particle experimentally discovered by Anderson with Dirac's hole particle (cfr. \citep[pp. 111-112]{Kragh:1990} and footnotes 104-105). In addition, Vladimir Fock posed serious challenges to Dirac's theory in a letter dated 12th February 1930: ``Firstly: Can the Pauli exclusion principle be applied to a continuous set of states (with continuous eigenvalues)? I always thought that it can only be applied to an innumerable set of states (discrete eigen-values), for in the formulation of this principle it seems to be necessary to numerate the states. For if the states may differ from one another by an infinitely small amount one can never say that all the states in a region of eigenvalues, however small, are filled up. In your theory, however, you apply the exclusion principle to all states, continuous as well as discontinuous'' quoted in \citep[p. 236]{Wright:2016}. To counter this objection, Dirac advanced the hypothesis to quantize space in infinitesimally small and impenetrable boxes in order to have discrete states of negative energy electrons, but then both this idea and Fock's critique were not further considered (cfr. \citep[p. 243]{Wright:2016}).}

Secondly, we have seen that Dirac employed methodological principles which can potentially contradict the observability doctrine. Discussing the proton interpretation of the holes, in fact, we underlined that he was led by an ontological belief concerning the unity of nature, Dirac being not only reticent to introduce additional particles to those already known, but also determined to show the metaphysical inter-dependence of electrons and protons. This principle can be understood as a sort of reductionist claim according to which nature is composed by very few essential elements. Interestingly, as we already noted, Dirac was not interested by the objections pointing out the empirical inadequacy of the ``hole=protons'' hypothesis, and this attitude is certainly in conflict with a positivistic methodology. Indeed, Dirac suddenly abandoned the unitary view when rigorous mathematical arguments showed that his theory entailed wrong consequences concerning the mass ratio between the electron and the proton. As said above, critiques pointing out empirical difficulties of the hole theory were of secondary importance for Dirac, and, again, this would be not acceptable by a physicist guided by the observability doctrine.

In the third place, another methodological principle adopted by Dirac is the principle of plenitude, as Kragh called it. According to its traditional formulation, everything that can be conceived as a metaphysical possibility, can be also an actuality in our world. In the context of theoretical physics -- which is where Dirac was moving -- such principle, with the due modifications, is often employed ``in the sense that entities are assumed to exist in nature as far as they are subject of mathematically consistent description and are not ruled out by the so-called principles of ``impotence'', or general statements that assert the impossibility of achieving something'' \citep[p. 271]{Kragh:1990}. Thus, according to this principle, if an entity $x$ is amenable of consistent mathematical treatment, its existence in nature cannot be in principle ruled out. Furthermore, to this regard it is important to stress that Dirac explicitly followed what \cite[pp. 222, 223]{Eddington:1923} called \emph{the Principle of Identification}; having defined a pure geometrical system Eddington affirms that it 
\begin{quote}
is intended to be descriptive of the relation-structure of the world. The relation-structure presents itself in our experience as a physical world consisting of \emph{space, time} and \emph{things}. The transition from the geometrical description to the physical description can only be made by identifying the tensors which measure physical quantities wit tensors occurring in the pure geometry; and we must proceed by inquiring first what experimental properties the physical tensor possesses, and then seeking a geometrical tensor which possesses these properties \emph{by virtue of mathematical identities}. If we can do this completely, we shall have constructed out of the primitive relation-structure a world of entities which behave in the same way and obey the same laws as the quantities recognised in physical experiments.
\end{quote}

\noindent Dirac, in fact, when explaining the new directions and advancements of theoretical physics claimed that:
\begin{quote}
[t]he most powerful method of advance that can be suggested at present is to employ all the resources of pure mathematics in attempts to perfect and generalise the mathematical formalism that forms the existing basis of theoretical physics, and \emph{after} each success in this direction, to try to interpret the new mathematical features in terms of physical entities (by a process like Eddington's Principle of Identification) \citep[p. 60]{Dirac:1931}.
\end{quote} 

Dirac made use of the plenitude principle advancing the anti-electron hypothesis: although anti-electrons were not yet observed, they were nonetheless a consistent solution of the DE, therefore, a realist interpretation of such particles was possible. Alternatively stated, in virtue of the plenitude principle, since the anti-electron hypothesis is mathematically consistent, such entities can represent physically real particles. Interestingly, Dirac used several times the plenitude principle in his works, for instance Kragh mentions Dirac's hypothesis concerning the existence of monopoles, which are not formally excluded by quantum theory. To this regard, Dirac clearly wrote that ``one of the elementary rules of nature is that, in the absence of law prohibiting an event or phenomenon it is bound to occur with some degree or probability. To put it simply and crudely: Anything that can happen does happen. Hence physicists must assume that the magnetic monopole exists unless they can find a law barring its existence'', (quoted in \cite{Kragh:1990}, p. 272). We will see with the second case study, the plenitude principle is again at work for Dirac tried to reintroduce a quantum mechanical ether. 

To conclude this section, we point out that in order to understand Dirac's methodology, it is crucial to underline that the principle of plenitude reflects a general rule often followed by the British physicist, i.e to deduce physics from mathematical considerations -- as shown by the above quotations --, meaning that physical conclusions are draw starting from the mathematical structure of a given theory. It is straightforward to note the tension and the contradiction between such methodology and the observability doctrine.  Furthermore, given the prominent importance given by Dirac to mathematics, it is not a surprise that the observability principle succumbed several times to other more important methodological lines. This fact is sufficient to show, in our opinion, that Dirac cannot be considered in any meaningful way of the term an empiricist or a positivist, despite his rhetorical defence of such philosophies. 

\subsection{Case study II: The Ether and Quantum Mechanics}

The second case study concerns Dirac's theory of electrodynamics; what is philosophically interesting for our discussion is that in this context Dirac explicitly reintroduced a quantum mechanical ether, a notion which has been abandoned at the beginning of the XX century following Einstein's theory of special relativity. As we will see, Dirac's ether is a formally unobservable theoretical entity. Thus, it provides a further example showing how Dirac's methodology was not guided by the observability doctrine to which he has often been associated. 
\vspace{2mm}

The British physicist was not satisfied with the way in which quantum electrodynamics developed during the thirties, since that theory had to face the well-known problems entailed by divergences and infinities. Dirac, as usual, tried to solve them by taking an unusual way; in order to tackle these issues, in fact, he thought it would have been necessary to start from a well-defined \emph{classical} theory of electrodynamics, i.e. a classical theory in which the central problem of the self-interaction of electron disappears:
\begin{quote}
We are now faced with the difficulty that, if we accept Maxwell's theory, the field in the immediate neighbourhood of the electron has an infinite mass. This difficulty has recently received much prominence in quantum mechanics (which uses a point model of the electron), where it appears as a divergence in the solution of the equations that describe the interaction of an electron with an electromagnetic field and prevents one from applying quantum mechanics to high-energy radiative processes. One may think that this difficulty will be solved only by a better understanding of the structure of the electron according to quantum laws. However, it seems more reasonable to suppose that the electron is too simple a thing for the question of the laws governing its structure to arise, and thus quantum mechanics should not be needed for the solution of the difficulty. \emph{Some new physical idea is now required, an idea which should be intelligible both in the classical theory and in the quantum theory, and our easiest path of approach to it s to keep within the confines of the classical theory} \citep[pp. 148-149, our italics]{Dirac:1938}.
\end{quote}
Such a theory was initially formulated in \cite{Dirac:1938} and then he continued to develop it in a number of publications appeared in the fifties. 

Hence, the British physicist proposed to start from a new classical theory based on more sound foundations in order to be able to improve the quantized version; this example shows the interplay between classical and quantum theories typical of Dirac's methodology, an attitude that many scholars labeled the reverse correspondence principle; we will give more details about these issues in the next section.\footnote{This route was taken before Dirac also by Adriaan Fokker, a collaborator of Lorentz, who was pushed by the pertinent problems of quantum electrodynamics to improve the classical theory. For details see \citep[p. 191]{Kragh:1990}.}

According to the classical theory of the electron as developed by Lorentz, Poincar\'e, and many other physicists, the electron was thought to be a spherical particle of finite size, or more precisely, a spherical distribution of electricity. In this context, if an electron is moving in an external magnetic field, it is subject to the Lorentz force. However, it also is subject to a self-force created by the field produced by the electron itself. This fact clearly was a source of problems, since physicists thought that the electron would be unstable in virtue of this self-force. In more detail, Lorentz showed that the electron's self-force can be written as follows:
\begin{align}
\label{self}
\vec{F}_{self}=-\alpha\frac{e^2}{ac^2}\dot{\vec{v}}+\frac{2}{3}\frac{e^2}{c^3}\ddot{\vec{v}} -\beta\frac{e^2a}{c^4}\dddot{\vec{v}}+\dots
\end{align}

\noindent where in this equation $\alpha, \beta$ are coefficients depending on the electron's structure, and the dots on $\vec{v}$ represent as usual the differentiation with respect to time. The self-force problem can be stated taking into account a point electron, i.e. an electron in which the electron's radius is zero ($a=0$). In this case the third and the higher terms vanish, but the first becomes infinite, on the other hand ``if $1/a$ is kept finite, the equation of motion contains not only an acceleration term but also derivatives of the acceleration to all higher orders'' \citep[p. 190]{Kragh:1990}. Against this background, Dirac's aim was to find the correct mathematical equations to model the electron's behavior with respect to the set of available evidence obtained from experiments.\footnote{In a passage taken from \citep[p. 149]{Dirac:1938} he wrote that ``[t]he scheme must be mathematically well-defined and self-consistent, and in agreement with well-established principles, such as the principle of relativity and the conservation of energy and momentum. Provided these conditions are satisfied, it should not be considered an objection to the theory that it is not based on a model conforming to current physical ideas''.}

Dirac solved the self-interaction problem starting from the usual Maxwell's equations, defining then the field quantities in terms of potentials. What is interesting for us is that he took into consideration not only retarded fields but also \emph{advanced fields}, which generally were considered unphysical solutions. Dirac's idea was to propose a symmetrical role between them, so that using both retarded and advanced field he was able to avoid the divergent term $v$ in his equations (cfr. for technical details \cite{Kragh:1990}, Chapter 9). Dirac, furthermore, developed Lorentz's classical theory arriving at an equation where neither infinity terms, nor structural dependencies were involved. Such theory is today known as Lorentz-Dirac theory. 

The difficult part of the work was, needless to say, its extension to quantum electrodynamics. As already stressed, Dirac aimed to remove the infinities of quantum electrodynamics starting from a classical theory in which such infinities do not occur. However, although he actually arrived at a classical theory without infinities, he was not able to extend this feature to the quantized theory; this fact is clearly expressed in a letter to Bohr dated 5th December 1938:
\begin{quote}
I spent the whole term working on the quantization of my classical electron theory. The first problem is to express the classical equations in Hamiltonian form. [...] With the classical theory in Hamiltonian form it is merely a mechanical matter to go over to the quantum theory. I have not yet satisfied myself, however, that the resulting quantum theory has no infinities. From the closeness of the analogy between classical and quantum theory one would expect that any classical theory from which the infinities have been eliminated would go over into a quantum theory without infinities (quoted in \cite{Kragh:1990}, pp. 195-196).
\end{quote}

Unfortunately, Dirac did not succeed in obtaining a quantum electrodynamics free of infinite terms. In the fifties, he continued with his work on classical electrodynamics still motivated by the same belief according to which, since there is a structural similarity and continuity between classical and quantum mechanics\footnote{The formal analogy between classical and quantum mechanics has been analyzed by Dirac in detail in his \emph{Principle of Quantum Mechanics}, and this issue was recurrent in Dirac's career; to this regard the reader may refer also to \cite{Dirac:1933} and \cite{Dirac:1945aa}.}, it would make sense to solve the problems of quantum theory putting the classical theory on firm mathematical grounds. In his work during the fifties, Dirac considered not individual electrons as the elementary blocks of his theoretical building, but rather a continuous stream of electricity: electrons as point particles would have then be considered \emph{effects} of quantum electrodynamics. It is in this context that Dirac re-introduced into physics the notion of an universal ether, which, he thought, could have been made compatible with the special theory of relativity. 

In the first place, Dirac reconstructed the argument against the ether that led to the elimination of this notion. Let us consider a perfect vacuum, i.e. a region of spacetime where there is no matter and no fields; according to the principle of relativity, in this region all the possible directions within the light-cone must be all equivalent. However, if an ether would exist at each space-time point of the region under consideration, it would have moved with a definite velocity (obviously less than $c$, otherwise it would have violated another axiom of special relativity). The latter would have consequently preferred one of the possible directions among those possible in the light-cone. Therefore, one obtains a contradiction with the relativity principle, since the ether would break the equivalence among the directions in the vacuum. In the second place, the ether's velocity could not be measured, so that, according to the positivistic attitude surrounding the special theory of relativity, something that could not possibly be observed should not be admitted as existing; thus, the notion of the ether was soon dismissed after the appearance of special relativity.

In order to show how such arguments did not definitely rule out the notion of an ether, Dirac suggested to take also the knowledge derived from quantum mechanics into account, and apply it to the ether:
\begin{quote}
The velocity of the ether, like other physical variables, is subject to uncertainty relations. For a particular physical state the velocity of the ether at a certain point of space-time will not usually be a well-defined quantity, but will be distributed over various possible values according to a probability law obtained by taking the square of the modulus of a wave function. We may set up a wave function which makes all values for the velocity of the ether equally probable. Such a wave function may well represent the perfect vacuum state in accordance with the principle of relativity. \citep[p. 605]{Dirac:1951}.
\end{quote}

Hence, if one characterizes the ether as a physical quantity with a state of motion, there are arguments from QM that enable one to reintroduce it among the realm of the acceptable theoretical entities. As noted by \citep[pp. 201-202]{Kragh:1990}, if the ether is subject to the laws of quantum theory, its ``velocity would be distributed over various possible values according to some probability law. The principle of relativity indeed forbade that there be any preferred direction of spacetime, which is a perfect vacuum, but this requirement could be reconciled with the ether hypothesis if one assumed that in a vacuum all velocities of the ether would be equally probable and distributed in a Lorentz-invariant way''.

The ether, then, was identified with the velocity field of the streams of electricity, giving it a realistic interpretation and a dignity as a proper physical entity. Such a quantum mechanical ether, however, was a weird kind of field being substantially a set of potentialities defined at certain space-time points; according to Dirac's theory, while the ontology of the theory was about the stream of electricity, the ether filled the vacuum. 

In a successive paper \citep{Dirac:1954}, Dirac restated his arguments characterizing the ether as a ``light and tenuous'' form of matter. Since the quantum indeterminacy applies to very small and light objects, Dirac argued, then it follows that the ether ``must be strongly affected by the principle of indeterminacy'' (p. 145). Therefore, at each space-time point the ether must have indeterminate values for its positions and momenta -- one among those permissible -- in order to not generate contradictions with relativity. 

It is clear from this brief presentation of Dirac's ideas concerning quantum electrodynamics that Eddington's principle of identification was implemented also in this case: this new ``quantum'' ether has been considered a serious physical possibility by Dirac since he was able to provide an argument (i) against its elimination, and (ii) its coherence with special relativity in the context of his new theory of electrodynamics. As in the case of the Dirac sea, the ether has been introduced as a possible theoretical entity referring to something physically existing. Furthermore, such a notion is inherently characterized as formally non-observable, since it is subject to Heisenberg's position-momentum uncertainty relation, which is taken by Dirac to forbid the simultaneous measurement -- and definition -- of position and velocity of quantum objects, and consequently of the quantum mechanical ether. Therefore, we can conclude that also in this case the observability doctrine has not been taken into consideration by Dirac in his reflections concerning the problems of infinities in quantum electrodynamics. 
\vspace{5mm}

Let us conclude this section stressing that Dirac himself unambiguously stated that unobservable theoretical entities cannot be dispensed with in the structure and vocabulary of physical theories, contrary to the fundamental tenets of the observability doctrine:
\begin{quote}
there must be unobservable quantities coming into the theory and the hard thing is to find what these unobservable quantities are \citep[p. 759]{Dirac:1973}.
\end{quote}

Finally, it is fair to claim that the observability principle -- in both the characterizations we have given above, the ontological and the causal -- ``did not affect his scientific work. In fact, he [Dirac] did not hesitate to propose quantities that seemed to have only the slightest connection to observables'' \citep[p. 264]{Kragh:1990}.

\section{Dirac and the Copenhagen Interpretations of Quantum Theory}
\label{Cop}

After having shown that the observability doctrine was endorsed rhetorically by Dirac in several places, but poorly considered in his own work, in this section we argue that he should be disentangled from the supporters of the Copenhagen interpretation of quantum theory. More specifically, in what follows we will show that often Dirac's views on quantum theory diverged remarkably -- and in an irreconcilable way -- from those of Bohr and Heisenberg. 

Let us underline in the first place that many scholars argued that it is disputable whether such interpretation exists, given that Bohr and Heisenberg had different -- often conflicting -- views concerning the physical and philosophical content of quantum theory.\footnote{The reader may refer to \cite{Howard:2004aa} and \cite{Jaeger:2009aa}, Chapter 3. Pauli is also often associated with the Copenhagen interpretation, but even his views differed substantially from those of Bohr. To support this claim it is sufficient to consider the positivistic attitude he had in the twenties, or the subjective interpretation of the quantum mechanical wave function; these that were not shared by the Danish physicist (cfr. the above mentioned references).} In particular, among other things, they disagreed about whether or not the wave function collapses in measurement situations: according to Bohr there was no collapse of the $\psi$ function -- being entanglement and complementarity the real novelties brought about by quantum mechanics with respect to classical physics --, whereas for Heisenberg, the observer induced quantum jumps are the primary innovation of quantum theory. Notably, to this regard \cite{Howard:2004aa} claimed that the so-called ``Copenhagen interpretation'' was Heisenberg's postwar invention: 
\begin{quote}
What was new in 1955 was Heisenberg's dubbing his amalgam of ideas the ``Copenhagen interpretation'', but having so dubbed it, Heisenberg regularly reinforced the invention of a unitary Copenhagen point of view
and posed as its chief spokesperson[...]. It helps to recall Heisenberg's situation in 1955, especially the fact that the person who was Bohr's favorite in the 1920s had become a moral exile from the Copenhagen inner circle in the postwar period, mainly because of the bitter rupture in Heisenberg's relationship with Bohr during his ill-fated visit to Copenhagen in September 1941 after taking over the leadership of the German atomic bomb project [...]. What better way for a proud and once ambitious Heisenberg to reclaim membership in the Copenhagen family than by making himself the voice of the Copenhagen interpretation? \citep[p. 677]{Howard:2004aa}.
\end{quote}

Interestingly for our purposes, if one considers the Copenhagen interpretation of quantum mechanics the expression of Bohr's views, it is straightforward to notice that under many respects Dirac's formulation of QM diverges remarkably from it. In this section, in fact, we will concentrate on the diverse conceptions (i) of quantum measurements, and (ii) of the correspondence principle that Bohr and Dirac had. Similarly, if one considers the Copenhagen interpretation as the expression of Heisenberg's perspectives on foundational issues about quantum theory, it is equally straightforward to show the remarkable differences with respect to Dirac's ideas. Not only the British physicist did not embraced in practice the young Heisenberg's observability doctrine as shown in the previous section, but also Dirac never shared the subjectivist interpretation of the quantum mechanical wave function endorsed by the later Heisenberg (cfr. \cite{Heisenberg:1958aa}, pp. 99-100). Furthermore, as we will see in the remainder of this section, they held opposite views about the inter-theoretic relations between classical and quantum mechanics (cfr. \cite{Bokulich:2004} on this point), a disagreement which is reflected in their incompatible methodologies.

Thus, given the peculiarity and originality of Dirac's ideas about the interpretation of quantum theory, we shall conclude that it is not quite correct to include him among the proponents of the Copenhagen view -- in both formulations given above -- despite the influence that Bohr and Heisenberg had on his work on quantum theory.

\subsection{Collapse or Not Collapse?}

According to the usual presentations of quantum mechanics, when a wave function undergoes a quantum measurement, it collapses into one of the admissible eigenstates of the measured operator. It is exactly the interaction between observed system and experimental device that causes the suppression of the Schr\"odinger equation and the consequent projection of the $\psi$ function. However, such treatment of quantum measurements was not endorsed by one of the founding fathers of quantum theory. Niels Bohr, in fact, never introduced explicitly the projection postulate in his works on quantum mechanics, as \citep[p. 672]{Howard:2004aa} explicitly stated: ``Bohr never endorsed a disturbance analysis of measurement [...]'', he ``always criticized Heisenberg for promoting the disturbance analysis, arguing that while indeterminacy implies limitations on measurability, it is grounded in limitations on definability''. Contrary to Heisenberg's ideas, according to the Danish physicist, the novelties introduced by quantum measurements -- and more generally by quantum mechanics -- are non-separability, entanglement and complementarity, as stated a few lines above.\footnote{For details concerning Bohr's philosophy of quantum mechanics the reader may refer to \cite{Howard:1994}, \cite{Howard:2004aa} and \cite{Jaeger:2009aa}, pp. 124 - 136.} To this specific regard, Bohr himself wrote that
\begin{quote}
the quantum postulate implies that any observation of atomic phenomena will involve an interaction with an agency of observation not to be neglected. Accordingly, an independent reality in the ordinary physical sense can neither be ascribed to the phenomena nor to the agencies of observation. After all, the concept of observation is in so far arbitrary as it depends on which objects are included in the system to be observed (\cite{Bohr:1934aa}, pp. 54-55).
\end{quote}

In this passage it is notably claimed that in the context of quantum theory -- contrary to the classical case --, it is not possible to assign an independent reality to the observed system and the measuring device, once they have interacted; using a technical jargon, they form what is now called an entangled pair. Therefore, the states of the observed system and that of the measuring device show a mutual dependence, and they cannot be written as they were separable, i.e. independent states as they were before the measurement.

To this specific regard, Bohr always emphasized that quantum phenomena -- that he generally defined as the observation of a certain quantity obtained under particular circumstances, i.e. with specific experimental arrangements, which play a crucial role in Bohr's theory of measurements  -- involve a mutual interconnection between the system that has been observed and the whole experimental situation used to measure it. Consequently, properties of quantum objects strictly depend on the interactions with the devices employed in measurement situations; thus, changing the experimental set-up will necessarily affect the nature of quantum systems:
\begin{quote}
The unambiguous account of proper quantum phenomena must, in principle, include a description of all relevant features of the experimental arrangement $[\dots]$. In the case of quantum phenomena, the unlimited divisibility of events implied in such an account is, in principle, excluded by the requirement to specify the experimental conditions. Indeed, the feature of wholeness typical of proper quantum phenomena finds its logical expression in the circumstance that any attempt at a well-defined subdivision would demand a change in the experimental arrangement incompatible with the definition of the phenomena under investigation. (\cite{Bohr:1963aa}, p. 3)
\end{quote}
 
Another central tenet of Bohr that comes out in this citation is the \emph{complementarity principle}. Although in quantum mechanics knowledge of physical systems is obtainable uniquely through measurements, there are nonetheless pieces of information about their properties that cannot be obtained simultaneously given the \emph{incompatibility of experimental protocols} needed to observe them, so that they cannot be represented by a unique quantum state of the examined system. For instance, according to quantum mechanics, it is not possible to obtain well-defined values for the position and the momentum of quantum particles in a single observation -- although one can easily measure these observables individually --, given the incompatibility of experimental procedures needed to observe them. Thus, the information obtainable by incompatible experiments is complementary.\footnote{Other classical examples of complementary properties of quantum systems are the wave-particle duality, or the spin of particles along different axis.} 

To this regard, (\cite{Stapp:2009aa}, p. 113) claims that ``any preparation protocol that is maximally complete, in the sense that all the procedures are mutually compatible and are such that no further procedure can add any more information, can be represented by a quantum state, and that state represents in a mathematical form all the conceivable knowledge about the object that experiments can reveal to us''. Thus, since for Bohr quantum states represent the complete description of physical systems (cfr. \citep[125]{Jaeger:2009aa}), it is clear (i) that the nature of quantum objects essentially depends on experimental protocols and measuring devices, and (ii) that observations do not reveal any pre-existing values of properties attributed to quantum systems. Thus, in virtue of the practical impossibility to experimentally show the complementary features of quantum objects, and the definition that Bohr gave to quantum phenomena, it follows that we cannot speak about the properties of quantum objects in isolation, so that one concludes that they have indeterminate features in non-measurement situations. 

Furthermore, Bohr thought that the quantum formalism cannot be applied to experimental devices, being strongly convinced that one should describe them \emph{classically}. More precisely, it is worth noting that Bohr strongly emphasized not only that the results of quantum measurements are necessarily expressed in terms of arrangements of macroscopic objects -- the only physical bodies that we can directly experience --, but also that the experimental procedure must be controllable and \emph{communicable} in order to provide an objective description of quantum phenomena:
\begin{quote}
it is decisive to recognize that, however far the phenomena transcend the scope of classical physical explanation, the account of all evidence must be expressed in classical terms \citep[p. 39]{Bohr:1958aa}.
\end{quote}
\noindent Hence, making communicability a necessary requirement for objectivity, classical, everyday concepts cannot be avoided in order to have some knowledge of quantum systems, since we have to ascribe ``definite properties to individual objects, a mode of description inherent in ordinary language and definitive of ``classical'' physics'' \citep[p. 674]{Howard:2004aa}.

Taking into account, instead, Dirac's formulation of quantum theory, it is straightforward to see that it provides a remarkably different account of experimental situations. In the first place, 
the primary difference is that he gives a completely formal treatment of measurements, contrary to the more qualitative descriptions of Bohr.
It is well-known, in fact, that Dirac defined a quantum system as described by a state vector $|\psi\rangle$, which is an element of a complex vector space called Hilbert space $\mathcal{H}$, providing a \emph{complete} specification of its properties which are represented by positive, linear Hermitian operators $\mathcal{A}$, acting on $\mathcal{H}$. According to Dirac's formulation of quantum mechanics, given a measurable quantity $A$, its possible values are the eigenvalues (real numbers) of the associated operator $\mathcal{A}$, whereas possible states in which a system may be found after a measurement of $A$ are represented by the eigenvectors of $\mathcal{A}$. Implicitly this defines the eigenvalue-eigenstate link, a core tenet of his approach to QM:
\begin{quote}
The expression that an observable `has a particular value' for a particular state is permissible in quantum mechanics in the special case when a measurement of the observable is certain to lead to the particular value, so that the state is in an eigenstate of the observable $[\dots]$. In the general case we cannot speak of an observable having a value for a particular state, but we can speak of its having an average value for the state. We can go further and speak of the probability of its having any specified value for the state, meaning the probability of this specified value being obtained when one makes a measurement of the observable. (\cite{Dirac:1947}, p. 253)
\end{quote}
In this quotation, the probabilistic and statistical character of quantum theory is clearly evident. This probability, however, refers to an inherent feature of the world, since only in measurement situations a specific value of the measured observable is obtained. Furthermore, according to the British physicist, the interactions occurring in measurement situations cause a stochastic ``jump'' of the wave function, a projection of $|\psi\rangle$ onto a possible eigenstate of the observed operator. This is exactly what makes QM inherently probabilistic for Dirac. He viewed these jumps as ``unavoidable disturbance'' of quantum systems in measurement situations:
\begin{quote}
When we measure a real dynamical variable $\xi$, belonging to the eigenvalue $\xi'$, the disturbance involved in the act of measurement causes a jump in the state of the dynamical system. From physical continuity, if we make a second measurement of the same dynamical variable immediately after the first, the result of the second measurement must be the same as the first. Thus after the first measurement has been made, there is no indeterminacy in the result of the second. Hence after the first measurement is made, the system is in an eigenstate of the dynamical variable $\xi$, [$\dots$]. In this way, we see that a measurement always causes the system to jump into an eigenstate of the dynamical variable that is being measured, the eigenvalue this eigenstate belongs to being equal to the result of the first measurement \citep[p. 36]{Dirac:1947}.
\end{quote}
Again, before the first measurement, the system's state is generally inherently indeterminate, since its properties depend strictly upon the act of observation. 

According to Dirac's views the remarkable novelties introduced by quantum theory are on the one hand, the non-commutative algebraic structure of its formalism, given the notable consequences it implies, and on the other hand, the stochastic collapses of the wave function, which in this approach to QM are actual physical processes occurring in space -- whereas for Bohr such random jumps did not play any substantial part in measurement situations. Furthermore, in Dirac's formulation of quantum theory one does not find claims concerning the objectivity of quantum measurements in terms of communicability of observational outcomes that are typical of Bohr's perspective.

In the second place, it is crucial for our argument to emphasize that also Bohr's principle of complementarity does not figure in Dirac's formulation of quantum mechanics. Rather than introducing a kinematical and dynamical complementarity in quantum theory associated with experimental protocols, he provided rigorous algebraic explanations why non-commuting observables, as for instance position and momentum, the particles' spin along different axis, etc., cannot be simultaneously measured, i.e. these operators do not share a common basis of eigenstates. Consequently, it follows that in Dirac's theory experimental protocols do not play any decisive role in determining the properties of a quantum object. Hence, Bohr's complementarity seems to play only a marginal role. 

\subsection{The Reciprocal Correspondence Principle}

Another fundamental tenet associated with the Copenhagen interpretation is Bohr's correspondence principle. In a nutshell, it affirms that classical mechanics is obtained -- through limiting procedures -- from quantum theory, furthermore as \citep[p. 126]{Falkenburg:2009} claims, ``Bohr employed the principle in order to establish inter-theoretical relations between the classical theory of radiation and the quantum theory of atomic spectra. After the rise of quantum mechanics, he justified his complementarity view of quantum mechanics in terms of the correspondence between mutually exclusive quantum phenomena on the one hand and the classical concepts of wave or particle [...] on the other hand''. Speaking about correspondence principle and inter-theoretic relations between classical and quantum mechanics, it is worth mentioning also Heisenberg's views. He thought that physical theories as classical Newtonian mechanics, thermodynamics, Maxwell's electromagnetic theory, special relativity or quantum mechanics, having reached a non-contradictory, definite system of axioms, definitions and laws, and being able to provide explanations for a notable set of phenomena, were \emph{closed}, in the sense that elements of these theories ``exhibit a tight interconnectedness prohibiting any further modifications or improvements'' \citep[p. 378]{Bokulich:2004}. Interestingly, Heisenberg considered these frameworks complete, accurate, and correct in their domain of applications for all times, meaning that such theories will be not modified or ``called into question by any future developments of science''.\footnote{Further details on Heisenberg's conception of physical theories and inter-theoretic relations are given in \cite{Bokulich:2006}.} 
 
Contrary to both these views, not only the principle of complementarity plays a little role in Dirac's formulation of quantum theory as we have seen above, but also the correspondence principle has been reversed. Taking into account the case study concerning the introduction of the quantum mechanical ether, in fact, we explained that Dirac's strategy to remove the infinities from quantum electrodynamics was to reformulate the \emph{classical} theory, where such infinite terms were already present. Thus, he tried to solve puzzles of quantum theories by starting from an improvement of the classical framework. This example is sufficient to show the tension between Dirac's methodology with respect to both Bohr's correspondence principle and Heisenberg's view of closed theories. As correctly pointed out by \citep[p. 386]{Bokulich:2004}, ``[f]or Dirac neither quantum mechanics nor classical mechanics has reached its final form''. Moreover, he considered physical theories always as approximations, and therefore, in continuous development and progress.\footnote{To this regard in the second session of his interview with Kuhn and Wigner mentioned in Section \ref{Obs}, Dirac stated: ``I think it's very likely that all our equations are only approximate. Our present quantum theory is probably only an approximation to the improvement of the future. I feel that everything might be an approximation and this comes very largely from the engineering training''.} 

Using Bokulich's expression, Dirac viewed both classical and quantum mechanics as open theories, since both these frameworks are subject to modifications and interplay:
\begin{quote}
My own opinion is that we ought to search for a way of making fundamental changes not only in our present Quantum Mechanics, but actually in Classical Mechanics as well. Since Classical Mechanics and Quantum Mechanics are closely connected, I believe we may still learn from a further study of Classical Mechanics. In this point of view I differ from some theoretical physicists, in particular Bohr and Pauli (Dirac quoted in \cite{Bokulich:2004}, p. 389).
\end{quote}

In the remainder of this section, we will consider another example showing the peculiarity of Dirac's positions with respect to those held by Bohr and Heisenberg concerning inter-theoretic relations among classical and quantum theories. 

Following his open view of theories, in the essay \emph{On the Analogy Between Classical and Quantum Mechanics} \citep{Dirac:1945aa}, Dirac aimed to recover and define the notion of particle trajectory in the context of quantum mechanics, giving to the latter a more intuitive, visualizable account for physical phenomena and extending the analogy between these two theoretical frameworks.\footnote{Since Dirac looked at the features and concepts of the classical theory to modify and improve the quantum, Alisa Bokulich introduced the expression \emph{reciprocal correspondence principle} (cfr. \cite{Bokulich:2004}, Section 6).} According to him, classical and quantum theories were closely connected having a strong formal similarity; to this regard, in fact, Dirac stated that 
\begin{quote}
The value of classical analogy in the development of quantum mechanics depends on the fact that classical mechanics provides a valid description of dynamical systems under certain conditions, when the particles and bodies composing the systems are sufficiently massive for the disturbance accompanying an observation to be negligible. Classical mechanics must therefore be a limiting case of quantum mechanics. We should thus expect to find that important concepts in classical mechanics correspond to important concepts in quantum mechanics, and, from an understanding of the general nature of the analogy between classical and quantum mechanics, we may hope to get laws and theorems in quantum mechanics appearing as simple generalizations of well-known results in classical mechanics \citep[p. 84]{Dirac:1947}.
\end{quote}
\noindent Similarly, at the outset of \citep[p. 195]{Dirac:1945aa}, he declared that 
\begin{quote}
There are two forms in which quantum mechanics may be expressed, based on Heisenberg's matrices and Schr\"odinger's wave functions respectively. The second of these is not connected very directly with classical mechanics. The first is in close analogy with classical mechanics, as it may be obtained from classical mechanics simply by making the variables of classical mechanics into non-commuting quantities satisfying the correct commutation relations. [...] In the case when the non-commuting quantities are observables, one can set up a theory of functions of them of almost the same degree of generality as the usual functions of commuting variables and one can use this theory to make closer the analogy between classical and quantum mechanics.
\end{quote}

In this paper, Dirac developed a new theory of functions able to assign a probability for non-commuting observables to have well-defined values. Let us consider two generic non-commuting observables $\alpha, \beta$, and let be $f(a, b)$ a function of two real variables $a, b$ defined when $a, b$ are eigenvalues of $\alpha, \beta$ respectively. Dirac showed how to assign a meaning to the function $f(\alpha, \beta)$. In the first place, define $f(\alpha, b)$ a function of the observable $\alpha$, in which the variable $b$ appears as parameter, via the following equation
\begin{align*}
f(\alpha, b)|\alpha'\rangle=f(\alpha', b)|\alpha'\rangle,
\end{align*}
\noindent where $|\alpha'\rangle$ represents an eigenstate of $\alpha$ with eigenvalue $\alpha'$.\footnote{This eqution is valid for every eigenvalue of $\alpha$.} Similarly one defines $f(\alpha, \beta)$ via  
\begin{align*}
f(\alpha, \beta)|\beta'\rangle=f(\alpha', \beta')|\beta'\rangle,
\end{align*}
\noindent where $|\beta'\rangle$ represents an eigenstate of $\beta$ with eigenvalue $\beta'$. Such function determines the linear operator $f(\alpha, \beta)$. Therefore, we have finally defined a general function of the two non-commuting observables $\alpha$ and $\beta$. This definition can be extended to the general case involving any number of non-commuting operators. It must be noted, however, that such definition is associated to an order of these non-commuting observables: ``[t]he observables that one uses in practice in Heisenberg's form of quantum mechanics are the values of dynamical variables at particular times. They fall into a natural linear order, namely the order of the times to which they refer. Our theory now enables us to set up general functions of them, based on this order. The functions must not involve two observables referring to exactly the same time, unless they commute. Apart from this limitation, the power of forming functions that we now have is just as general as in the classical theory'' \citep[p. 196]{Dirac:1945aa}. Dirac associated these new functions a complex probability value, which should be interpreted as the meaning that for quantum observables $\alpha, \beta, \gamma, \dots$ are unlikely to have eigenvalues $\alpha', \beta', \gamma', \dots$. As a final step he used these probabilities to introduce quantum trajectories.
In more details, in this theory, a quantum trajectory is constructed from a series of transition amplitudes between pairs of adjacent points $x_i$ at different, successive times $t_i$.\footnote{For further details, the reader may refer to the original paper \cite{Dirac:1945aa}; interesting discussions about quantum trajectories in the context of Dirac's work and its connections with David Bohm are contained in \cite{Hiley:2018, Hiley:2019}.} Time can be so small allowing to divide the trajectory into infinitesimal segments: $\langle x'_t|x_{t_0}\rangle$ is then the \emph{probability amplitude} of a quantum particle traveling in space from point $x$ at time $t_0$ to point $x'$ at time $t>t_0$. The particle may travel through intermediate points as well (if $x, x'$ are not adjacent). In the general case (when a particle passes through intermediate points), one may write: 
\begin{align*}
\langle x'_t|x_{t_0}\rangle=\int \dots \int \langle x'_t|x_n\rangle dx_n\langle x_n|x_{n-1}\rangle dx_{n-1}\dots  dx_1\langle x_1|x_{t_0}\rangle
\end{align*}

\noindent where $\langle x_{i+1}|x_i\rangle$ is the \emph{propagator} of the particle being at $x_i$ at time $t_i$ and arriving at point $x_{i+1}$ at time $t_{i+1}$. 

Since this probability is in general a complex number, as already pointed out, Dirac claimed that this theory gets ``a formal probability for the trajectory of the system in quantum mechanics lying within certain limits. This enables us to speak of some trajectories being improbable and others being likely'' \citep[p. 197]{Dirac:1945aa}, and therefore it provides a more intuitive picture of the motion of quantum objects in space.\footnote{This paper, although not very much discussed in the philosophical literature, has influenced Feynman's work on the path integral formulation of quantum mechanics. For details the reader may refer to \citep[p. 367]{Feynman:1948} and \cite{Schweber:1994}, Chapter 8.} 
\vspace{5mm}

In sum, having underlined the contrasting views Dirac held with respect to many fundamental tenets of Bohr's and Heisenberg's perspectives on quantum mechanics, we conclude that it would be erroneous to claim that Dirac was a fervent supporter of the Copenhagen interpretation.

\section{Conclusion}
\label{conc}

Although Dirac's ideas on quantum theory have been heavily influenced by Bohr and Heisenberg, physicists with whom he had intense and long correspondence and personal relationships, in this essay we have argued on the one hand that Dirac should not be considered a scientist methodologically guided by the observability doctrine, despite his explicit defence of the latter in several places; on the other hand, we also claimed that his views concerning quantum theory should be disentangled from those of the Copenhagen interpretation. 

More specifically, after having introduced the tenets of the observability principle, we discussed two case studies showing how such principle played little role in Dirac's own works, since he postulated the existence of formally unobservable theoretical entities -- which are also causally inert -- as the infinite sea of electron and the quantum mechanical ether. Hence, we concluded that his support of the observability doctrine was substantially more rhetorical than practical. Furthermore, not only we showed how Dirac's views about the quantum theory of measurement come into conflict with the views of Bohr, but we also emphasized his different ideas towards the correspondence principle and the inter-theoretic relations between classical and quantum mechanics, claiming that his perspective on quantum theory -- and on physics and mathematics in general -- should be kept sharply separated from those held by Bohr and Heisenberg. 

In sum, reading Dirac's works one cannot but note how complex a personality he was, a scientist difficult to insert in predefined categories due to his creative and innovative approach to physics and mathematics; in a century in which the new quantum paradigm was developed and imposed itself, and classical ideas were seen as definitively surpassed, he viewed physical theories as essentially mathematical structures which are always subject to modifications, changes and improvements, without any pretension to achieve an ultimate description of reality, and without any dogmatic attitude towards the future progress of physics. 
\vspace{5mm}

\textbf{Acknowledgements}

Andrea Oldofredi is grateful for support from the Swiss National Science Foundation, grant no. 105212-175971; AO wishes also to thank Olga Sarno for helpful comments on previous drafts of this paper.

\clearpage
\bibliographystyle{apalike}
\bibliography{PhDthesis}
\end{document}